\documentclass{aa}

\usepackage{txfonts}
\usepackage{natbib}
\usepackage{graphicx}
\usepackage{multirow}

\usepackage{color}
\newcommand{\lbdd}{$\lambda/D_{\rm L}$}

\bibpunct{(}{)}{;}{a}{}{,}

\begin{document}

\title{On-sky performance of the QACITS pointing control technique with the Keck/NIRC2 vortex coronagraph}
\titlerunning{On-sky performance of the QACITS pointing control technique}

\author{
   E.~Huby\inst{1}\fnmsep\thanks{F.R.S.-FNRS Postdoctoral Researcher}
\and M.~Bottom\inst{2}
\and B.~Femenia\inst{3}
\and H.~Ngo\inst{4}
\and D.~Mawet\inst{2,5}
\and E.~Serabyn\inst{2}
\and O.~Absil\inst{1}\fnmsep\thanks{F.R.S.-FNRS Research Associate}
}
\institute{
\inst{1} Space sciences, Technologies, and Astrophysics Research (STAR) Institute, Universit\'e de Li\`ege, 19c All\'ee du Six Ao\^ut, 4000 Li\`ege, Belgium\\
\inst{2} Jet Propulsion Laboratory, California Institute of Technology, 4800 Oak Grove Drive, Pasadena, CA 91109, USA\\
\inst{3} W. M. Keck Observatory, 65-1120 Mamalahoa Hwy., Kamuela, HI 96743, USA\\
\inst{4} California Institute of Technology, Division of Geological and Planetary Sciences, 1200 E. California Blvd, Pasadena, CA 91125, USA\\
\inst{5} California Institute of Technology, Division of Physics, Mathematics and Astronomy, 1200 E. California Blvd, Pasadena, CA 91125, USA
}


\abstract 
{
A vortex coronagraph is now available for high contrast observations with the Keck/NIRC2 instrument at L band. The vortex coronagraph uses a vortex phase mask in a focal plane and a Lyot stop in a downstream pupil plane to provide high contrast at small angular separations from the observed host star.
} {
Reaching the optimal performance of the coronagraph requires fine control of the wavefront incident on the phase mask. In particular, centering errors can lead to significant stellar light leakage that degrades the contrast performance and prevents the observation of faint planetary companions around the observed stars. It is thus critical to correct for the possible slow drift of the star image from the phase mask center, generally due to mechanical flexures induced by temperature and/or gravity field variation, or to misalignment between the optics that rotate in pupil tracking mode.
} {
A control loop based on the QACITS algorithm for the vortex coronagraph has been developed and deployed for the Keck/NIRC2 instrument. This algorithm executes the entire observing sequence, including the calibration steps, initial centering of the star on the vortex center and stabilisation during the acquisition of science frames. 
}{
On-sky data show that the QACITS control loop stabilizes the position of the star image down to 2.4\,mas rms at a frequency of about 0.02\,Hz. However, the accuracy of the estimator is probably limited by a systematic error due to a misalignment of the Lyot stop with respect to the entrance pupil, estimated to be on the order of 4.5\,mas. A method to reduce the amplitude of this bias down to $1\,$mas is proposed.
}{
The QACITS control loop has been successfully implemented and provides a robust method to center and stabilize the star image on the vortex mask. In addition, QACITS ensures a repeatable pointing quality and significantly improves the observing efficiency compared to manual operations. It is now routinely used for vortex coronagraph observations at Keck/NIRC2, providing contrast and angular resolution capabilities suited for exoplanet and disk imaging.
}


\maketitle

\section{Introduction}

In June 2015, a new coronagraphic mode available with the Keck/NIRC2 instrument \citep{Serabyn2017} came online. The L band imager is now equipped with a vector vortex coronagraph based on an annular groove phase mask \citep[AGPM,][]{Mawet2005}. It consists of a circular subwavelength grating etched onto a diamond substrate \citep{VargasCatalan2016}, placed in an intermediate focal plane wheel and working in conjunction with a downstream Lyot stop. Combined with Adaptive Optics (AO) correction and advanced post-processing techniques \citep[e.g.][]{Soummer2012, Amara2012, GomezGonzalez2016}, this mode allows high-contrast imaging of planetary companions and circumstellar disks around stars at very small angular separations, with an effective inner working angle (IWA) of $~0\farcs12$. A 5-$\sigma$ sensitivity limit of 10\,mag at $0\farcs5$ for a star of magnitude K=5 has been demonstrated during commissioning \citep{Absil2016}. In addition, such an imager at L band benefits from advantageous atmospheric conditions and generally more favourable planet-to-star contrast ratio than at shorter wavelengths, especially in the case of young self-luminous giant exoplanets. As such, the vortex mode at Keck/NIRC2 provides a promising tool to directly image exoplanets on relatively compact orbits, thus complementing second generation high contrast instruments working in the near infrared, e.g. SPHERE \citep{Beuzit2008}, GPI \citep{Macintosh2014} or SCExAO \citep{Jovanovic2015}.

With the implementation of this new mode, the necessity of a fine pointing control is crucial for two reasons: operation efficiency and contrast performance. First, observing with a vortex coronagraph requires the star image to be initially centered onto the vortex mask, and if done manually, this process can prove to be challenging and time consuming. As illustrated in \cite{Huby2015, Huby2016} and recalled in the next section, this is particularly true in case of centrally obstructed telescopes, where counter-intuitive flux asymmetry can appear when the star image is very close to the center of the mask. This effect can lead to the misinterpretation of the offset direction needed to improve the centering of the star image and make the centering process long and/or not optimal. Repeatability and rapidity of the alignment were thus major incentives for the implementation of an automated pointing control system.

Additionally, reaching the optimal performance of the coronagraph requires a high quality wavefront. Although aberrations due to turbulence are largely removed by the highly efficient Keck AO system, routinely delivering Strehl ratios on the order of 85-90\% at L band, noncommon path wavefront errors inside the instrument are inevitable and the source of substantial loss of contrast performance. Indeed, reaching a small IWA with a focal-plane phase mask coronagraph comes at the price of a high sensitivity to low order aberrations. The most detrimental of there are generally tip and tilt aberrations due to mechanical flexures and/or optics rotating for pupil tracking, which are common on large telescopes. The control and correction of these aberrations are thus crucial and require the implementation of additional dedicated sensors. Here, we focus on low order aberration sensors, which are critical in particular for small IWA coronagraphs. These sensors usually work in conjunction with other techniques sensitive to high-order noncommon path aberrations, such as speckle nulling, electric field conjugation, phase diversity, Zernike wavefront sensing, or interferometric methods \citep[see][and references therein]{Bottom2017}. As a matter of fact, speckle nulling was recently implemented on Keck/NIRC2 for use with the vortex coronagraph \citep[][Bottom et al. in prep]{Bottom2016}.

Concerning low order aberration control, several solutions have been developed and implemented in current high contrast instruments. In order to catch most of the noncommon path errors, the extra sensor must be placed as close as possible to the coronagraphic mask. For instance, in the case of the Differential Tip-Tilt Sensor (DTTS) integrated to the coronograph of SPHERE \citep{Baudoz2010}, a beamsplitter placed right before the focal plane phase mask sends a few percent of the light towards a separate detector in order to monitor the jitter and drift of the star image. Another solution consists in using the light that is rejected by the coronagraph, i.e. that is stopped by the occulting focal plane mask \citep[Coronagraphic Low-Order Wave-Front Sensor, CLOWFS,][]{Guyon2009}, like in GPI \citep{Wallace2010}, or stopped by the diaphragm in the Lyot plane, in case of a phase mask \cite[Lyot-based Low-Order Wave-Front Sensor, LLOWFS,][]{Singh2014}. Only the LLOWFS system has been characterized on-sky, reporting pointing residuals of 0.23\,mas or 5.8$\times10^{-3}\,\lambda/D$ \citep{Singh2015}. However, it has to be noted that all these sensors are inherently not fully common path and require a specific optical layout, which can make them challenging to integrate to an existing system.

Another kind of solution is based on the sole analysis of the coronagraphic images. In this case, the sensor is fully common path, but the number of modes that can be corrected is currently reduced to tip and tilt. Such sensor has been first proposed and tested in laboratory for the four quadrant phase mask (FQPM) coronagraph \citep{Mas2012} in case of a non obstructed pupil, leading to an accuracy of 6.5$\times10^{-2}\,\lambda/D$ measured in laboratory. The principle was then extended to the case of the vortex coronagraph with a centrally obscured pupil \citep{Huby2015}. As a completely non-invasive method that does not require any setup modification, this technique called QACITS (Quadrant Analysis of Coronagraphic Images for Tip-tilt Sensing) thus appeared as one of the logical and most accessible solutions for implementation on the new vortex coronagraph of the Keck/NIRC2 instrument. This paper reports on the practical implementation and the on-sky performance of this pointing sensor. In the next section, the principle of this technique is recalled and the different functions performed by the controller implemented for the Keck/NIRC2 instrument are described. In Sect.~\ref{sec:calib}, the experimental calibration of the tip-tilt estimator is presented, and possible sources of bias are investigated. In Sect.~\ref{sec:perf}, the on-sky performance of the control loop is assessed and compared with results obtained without the QACITS controller. Lastly, in Sect.~\ref{sec:conclusion}, the status of the algorithm is discussed and further ongoing developments are described.

\section{Implementation of QACITS at Keck/NIRC2}
\label{sec:qacits_loop}

\begin{figure}
\centering
\includegraphics[height=4cm]{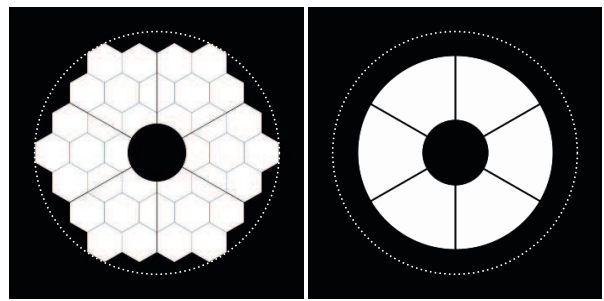}
\caption{Pupil masks used in the simulations of the Keck/NIRC2 instrument, as defined in Table~\ref{tab:pupil_keck}: entrance pupil on the \textit{left} and Lyot stop on the \textit{right}. The circumscribed circle in white dotted line has a diameter of 10.93\,m.}
\label{fig:keck_pupil}
\end{figure}

\begin{table}
\caption{Pupil configuration of the Keck telescope and Lyot stop mask used for the vortex coronagraph in NIRC2.}
\label{tab:pupil_keck}
\centering
\begin{tabular}{c|c}
\hline \hline
parameter & normalized value \\
\hline
Entrance pupil diam. & 1.00 \\
(circumscribed circle) & \\
Central obstruction diam. & 0.24 \\
Entrance pupil spider width & 0.0023 \\
Inter-segment gap & 2.7$\times10^{-4}$ \\
Lyot stop external diameter & 0.80 \\
Lyot stop internal diameter & 0.27 \\
Lyot stop spider width & 0.0061 \\
\hline
\end{tabular}
\end{table}

Throughout the paper, the tip-tilt amplitude is given either in milliarcseconds (mas) or in units of the angular resolution element defined as \lbdd, with $\lambda$ the wavelength and $D_{\rm L}$ the equivalent diameter of the Lyot stop, which defines the angular resolution in the final image. In practice, this diameter is equal to 8.7\,m and corresponds to 80\% of the entrance pupil diameter, defined here as the diameter of the circumscribed circle, i.e. 10.93\,m. The Lyot stop almost matches the inscribed circle in the hexagonal entrance pupil, which is 9\,m in diameter. The wavelength is taken as the central wavelength of the L band filter, i.e. $\lambda=3.776\,\mu$m, which gives \lbdd=89.3\,mas. The entrance pupil and Lyot stop shapes used in the simulations presented in this paper are shown in Fig.~\ref{fig:keck_pupil}. The latter corresponds to the incircle pupil mask available in the NIRC2 instrument. The pupil feature dimensions are reported in Table\,\ref{tab:pupil_keck}.

In this section, the principle of the QACITS estimator corresponding to the Keck telescope pupil is described, and the different operations performed by the controller implemented on Keck/NIRC2 are detailed.

\subsection{The QACITS estimator}

The QACITS estimator aims to measure the pointing errors affecting the beam incident on a coronagraphic phase mask directly from the analysis of the coronagraphic image shape \citep{Mas2012,Huby2015}. As a result, it probes aberrations that are fully common path with the coronagraphic mask. Given the flux level remaining after the coronagraph, aberrations induced downstream the mask have a negligible impact on the contrast performance. The QACITS estimator is based on the measurement of differential intensities, resulting from the integration and subtraction of the flux in the halves of the image, as in a quadrant position sensor, normalized by the total flux of the non coronagraphic point spread function (PSF).

\begin{figure}
\centering
\includegraphics[height=4cm]{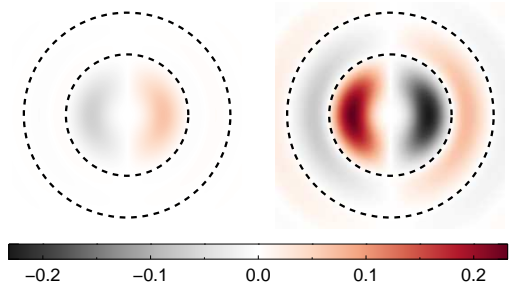} 
\caption{Images of the two asymmetric components appearing in the final coronagraphic image shape in presence of horizontal tip-tilt aberration. Positive tip-tilt amplitude induces a shift of the star image towards the right. \textit{Left}: $A_{\rm circ} \cos \psi$, contribution of the circular non-obstructed pupil. \textit{Right}: $A_{\rm obsc}\cos \psi$, contribution of the central obstruction. The circles delimiting the inner and outer regions have a radius of 1.6\,\lbdd and 2.7\,\lbdd. A comparison of the horizontal profiles of these images is displayed in Fig.\,6 of \cite{Huby2015}.}
\label{fig:asym_bessel_profiles}
\end{figure}

\cite{Huby2015} have previously shown that the electric field in the case of a centrally obstructed pupil can be described as the sum of two contributions related respectively to the circular unobstructed pupil and the central obstruction (added negatively in amplitude), leading to two distinct contributions in the flux asymmetry of the final image. These two components are expressed as combinations of Bessel functions modulated by a $\cos \psi$ term with $\psi$ the azimuthal angle, as it was established by Eq.~9 and~19 in \cite{Huby2015}. In short, for a tip-tilt amplitude $T$ applied along the $x$ axis, the final image $I$ can be described as the sum of a symmetric term and two asymmetric terms:
\begin{equation}
I \varpropto I_{\rm sym} +  T^3 \times A_{\rm circ} \cos \psi  + T \times A_{\rm obsc} \cos \psi,
\end{equation}
where $A_{\rm circ}$ and $A_{\rm obsc}$ are combinations of Bessel functions corresponding to the contributions of the circular pupil and central obstruction, respectively. The shape of the two asymmetric terms are displayed in Fig.~\ref{fig:asym_bessel_profiles}. They can be decomposed in two concentric regions, defined by the sign inversion of their amplitude. The boundaries between inner and outer regions are located at radii of 1.6\,\lbdd and 2.7\,\lbdd. The weight of each component in the final image is a function of the tip-tilt amplitude, making each term contribute differently to the asymmetry of the image in the two regions:

\begin{itemize}
\item the differential intensity due to the ideal circular pupil is 40 times stronger in the inner than the outer region of the image. In the final image, this contribution grows with the cube of the tip-tilt amplitude.
\item the amplitude of the differential intensity due to the central obstruction is about 3 times stronger in the inner than the outer region of the image, and they have opposite signs. This contribution varies linearly with the tip-tilt amplitude.
\end{itemize}

To complete the quantitative comparison, the differential intensity amplitude due to the circular aperture in the inner region (Fig.\,\ref{fig:asym_bessel_profiles}, \textit{left}) is about 4 times lower than the differential intensity amplitude due to the obstruction in the same region (\textit{right}).

\begin{figure}
\centering
\includegraphics[width=0.8\linewidth]{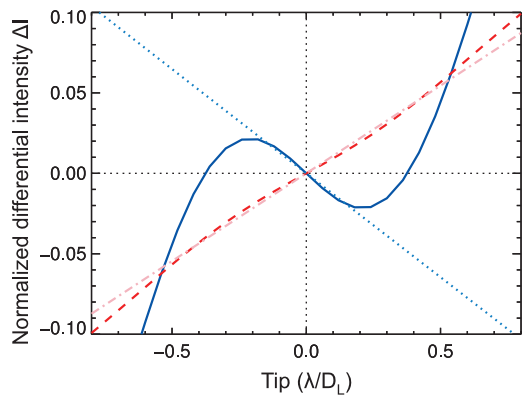} 
\caption{Differential intensity measured on simulated images with tip-tilt applied along the $x$ axis. The blue solid line corresponds to the flux integrated in the inner area only, while the red dashed line corresponds to the flux integrated in the outer area of the image. The light blue dotted and red dash-dotted lines are the linear approximations for the inner and outer measurements, respectively, that are valid in the small tip-tilt regime (i.e. $<0.15\,$\lbdd and $<0.5\,$\lbdd, respectively).}
\label{fig:ideal_model}
\end{figure}

\begin{figure}
\centering
\includegraphics[angle=270,width=.8\linewidth]{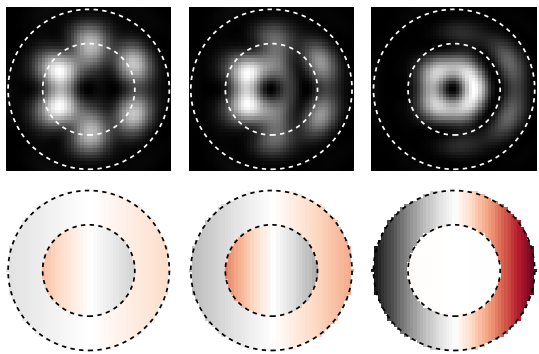} 
\caption{\textit{Top}: Simulated images for different tip-tilt amplitudes applied to the right along the horizontal axis, from left to right: 0.1\,\lbdd, 0.16\lbdd and 0.38\,\lbdd. Each displayed image is normalized by its maximal value. The dotted circles show the boundaries for the inner and outer areas. \textit{Bottom}: Color-coded representation of the flux asymmetry to emphasize the visual comparison of the asymmetry in the different images. The amplitude of the gradient is equal to the measured differential intensity and all images are shown with the same color scale.}
\label{fig:simulated_img}
\end{figure}

The analysis of the image in these two separate concentric regions allows the partial disentanglement of the two contributions. In the inner area, the two terms are present and have opposite signs, thus compensating each other and making the differential intensity measurement ambiguous, as can be seen in Fig.~\ref{fig:ideal_model}. This is also the reason why the observer can be misled, since the flux asymmetry in this region goes in the direction opposite to the actual offset of the star, as illustrated in Fig.~\ref{fig:simulated_img}. Moreover, for a tip-tilt amplitude of about 0.4\,\lbdd the central region looks like a symmetric doughnut, which can be additionally confusing (see Sect.~\ref{sec:pointing_stats}). In the outer area however, the linear contribution due to the central obstruction prevails, avoiding ambiguity in the measurement.

In practice, the QACITS estimator used at Keck/NIRC2 is based on the differential intensity measured in the outer area only, approximated by a linear model, which is valid in the small tip-tilt regime (i.e. $< 0.5\,$\lbdd, see Fig.~\ref{fig:ideal_model}). The linear approximation of the inner estimator is also monitored, since it can carry useful information (see Sect.~\ref{sec:effect_lyotshift}). Hereafter, the estimators based on the inner and outer parts of the image using the linear approximation will be designated as the inner and outer estimators, respectively. It should be noted that the shapes of the differential intensity curves depend on the telescope configuration. In particular, for a smaller central obstruction (as is the case for VLT/NaCo or LBT/LMIRCam with 14\% and 11\% central obstruction, respectively), the slopes of the linear model approximations would be lower.

\subsection{Closed-loop implementation at Keck/NIRC2}
\label{sec:keck_loop}

The QACITS algorithm has been implemented in IDL and takes care of the initial centering optimization, the pointing correction and the science acquisitions, making observations with the vortex mode user-friendly and highly time-efficient. Before calling the QACITS sequence, the observer only has to make sure that the star is roughly centered onto the vortex mask (within $1\,$\lbdd), a pointing requirement that is routinely achieved by the AO system once a reference position has been saved. The QACITS controller then operates in three steps:
\begin{itemize}
\item A calibration step, which takes an unsaturated image of the star far off the center of the vortex mask (off-axis PSF), and sky images. The position of the center of the vortex mask is identified by fitting the vortex center glow that is visible in sky images \citep{Absil2016}. The actual offset of the star image is then estimated by fitting a Gaussian profile to the off-axis PSF, and this offset is corrected by sending a telescope offset command in units of pixels. At the end of this step, the star is roughly centered onto the vortex phase mask (typically within a few 0.1\,\lbdd). This step usually requires $\sim$2\,min to be completed.
\item An optimization sequence, which consists of a few iterations of the QACITS loop that are run faster than the scientific acquisitions (using shorter integration time and smaller frame size to minimize NIRC2 overheads). By default, the alignment is considered optimized when the measured residual tip-tilt has an amplitude smaller than 0.1\,\lbdd. This criterion can be tuned in the QACITS parameters (the observer can require that several consecutive estimations fall below a chosen limit). With the default settings of $T_{\rm int}$ = 0.2\,s (integration time) and $N_{\rm coadd}$ = 10 (number of co-added images) for a frame width of 512~pixels (minimal value for the vortex center to be included in the sub-image), one iteration is 20\,s long. This sequence typically takes 1-2~min.
\item The science acquisition sequence, with the QACITS correction applied after each acquisition. With the typical settings of $T_{\rm int}$ = 0.5\,s and $N_{\rm coadd}$ = 50 for a full frame width of 1024~pixels, one science acquisition is 46\,s long (for 25\,s of actual integration time).
\end{itemize}

The correction algorithm applied during the science sequence consists of a proportional-integral controller, with proportional and integral gains $G_{\rm P} = 0.3$ and $G_{\rm I} = 0.1$, which have been tuned experimentally to ensure the stability of the loop. The loop is run at a frequency defined by the time needed for one acquisition, i.e about 0.02\,Hz, hence correcting for the slow drift (see Sect.~\ref{sec:slow_drift}). The QACITS calling sequence and parameters for Keck/NIRC2 are described in more details in \cite{Huby2016} and in a user manual available online\footnote{https://www2.keck.hawaii.edu/inst/nirc2/observing}.
The pixel size of the coronagraphic images is 10\,mas/pixel \citep{Service2016}, i.e. about 9 pixels per resolution element \lbdd. Since the PSF is well sampled, we have not investigated in detail the effect of low sampling on the QACITS estimator. This will be studied in the future, in particular for application on other instruments.

\section{Vortex mode characterization}
\label{sec:calib}

\subsection{On-sky calibration}
\label{sec:onsky_calib}

\begin{figure}
\centering
\includegraphics[width=\linewidth]{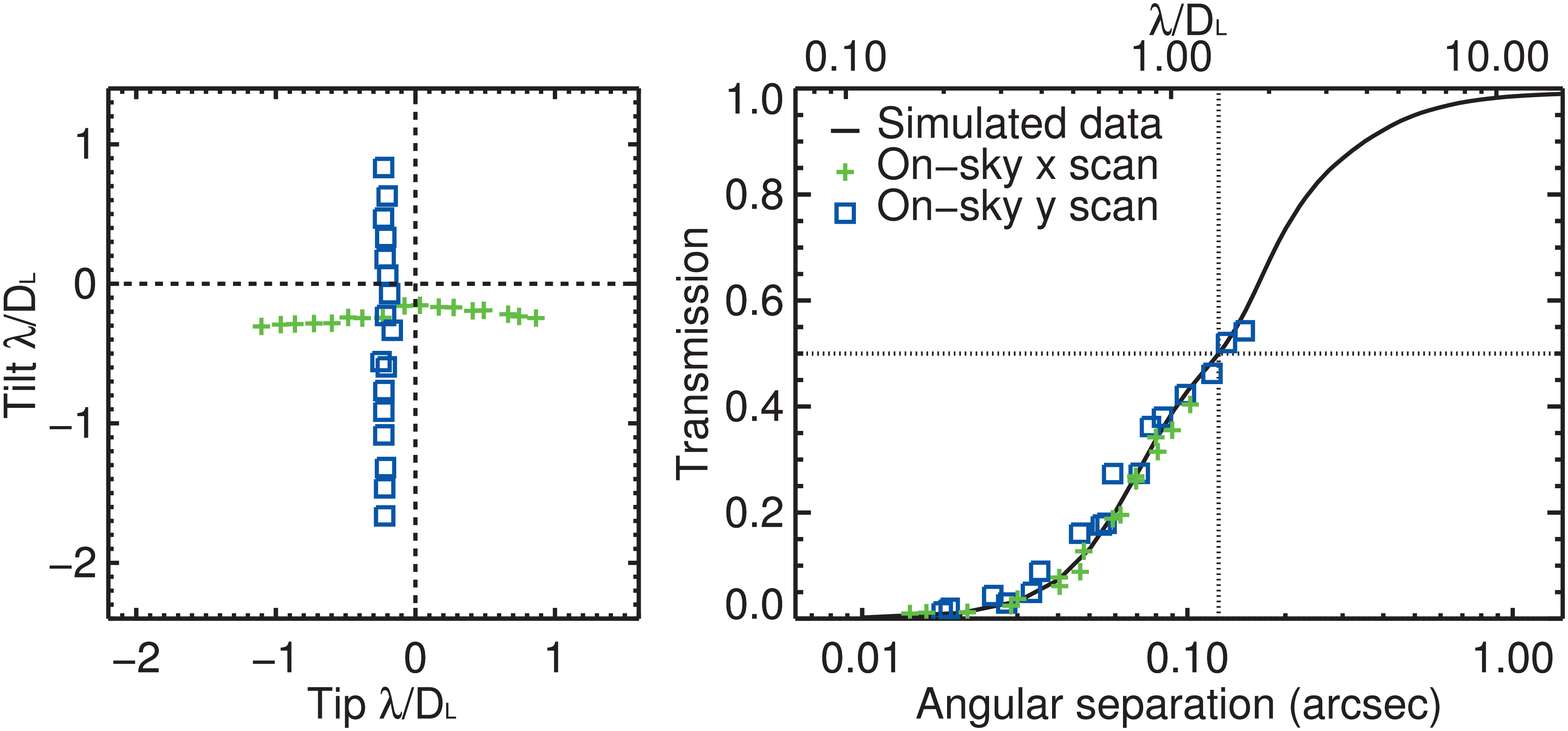} 
\caption{Tip-tilt scanning sequences carried out for characterizing the QACITS estimator behaviour. \textit{Left:} True tip-tilt values as monitored by the secondary component of the binary system, for the scanning sequences along the $x$ (green crosses) and $y$ (blue squares) axes. \textit{Right:} Comparison of the simulated peak transmission curve (flux integrated on a disk of diameter $1\,$\lbdd) with the experimental results measured from the on-sky data. The effective IWA is reached at a separation of 1.4\,\lbdd = 125\,mas (computed from the simulated curve).}
\label{fig:exp_true_tiptilt}
\end{figure}

The calibration of the model for the QACITS estimators has been performed on-sky. For that purpose, a wide binary system was observed, with the brightest component of the system centered on the vortex mask. The monitoring of the companion position provides a means to estimate the true position of the star image behind the mask. On October 29th, 2015, acquisitions were thus taken on the binary system of HD46780 ($L_{\rm primary}$=5.5, $L_{\rm secondary}$=7.2). The separation was 737.2\,mas at the date of the observation \cite[orbit parameters from][]{Heintz1993}. The long period of 118.9~years ensures that this separation does not vary during the time of observation.

\begin{figure*}
\centering
\includegraphics[width=15cm]{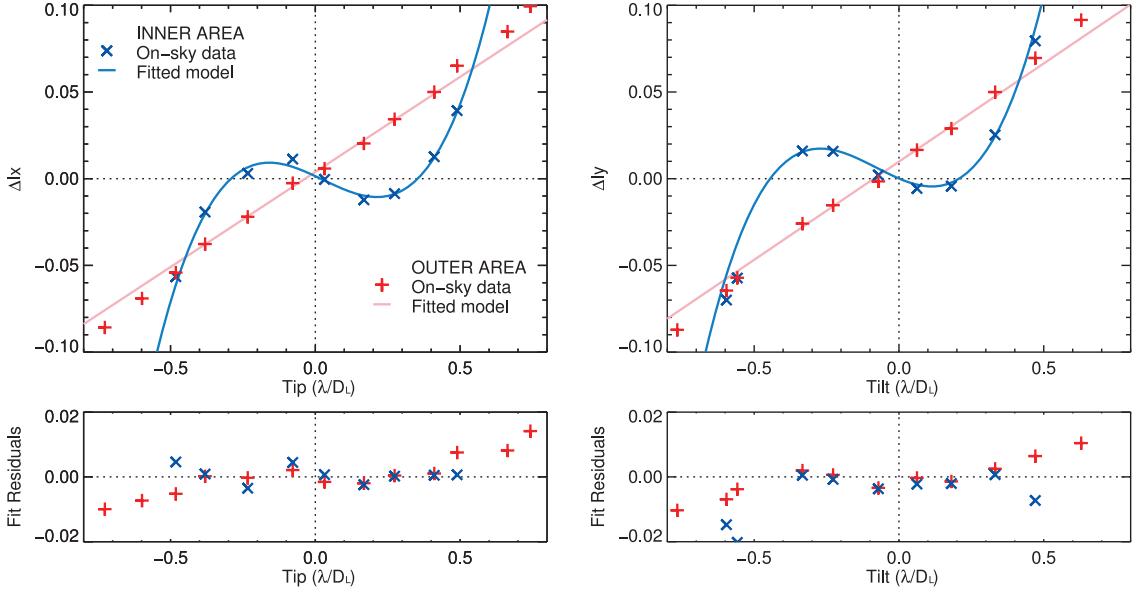} 
\caption{Experimental characterization of the QACITS model from on-sky data. Left and right plots correspond to the differential intensity measurements as a function of tip-tilt for the $x$ and $y$ directions, respectively. In both cases, the differential intensity was computed in the inner (blue crosses) and outer (red pluses) areas of the image. The best fit models in the least-squares sense are shown in solid lines (see Eq.~\ref{eq:model} for their definition), and the residuals are shown in the plots below.}
\label{fig:exp_calib1}
\end{figure*}

\begin{figure*}
\centering
\includegraphics[width=15cm]{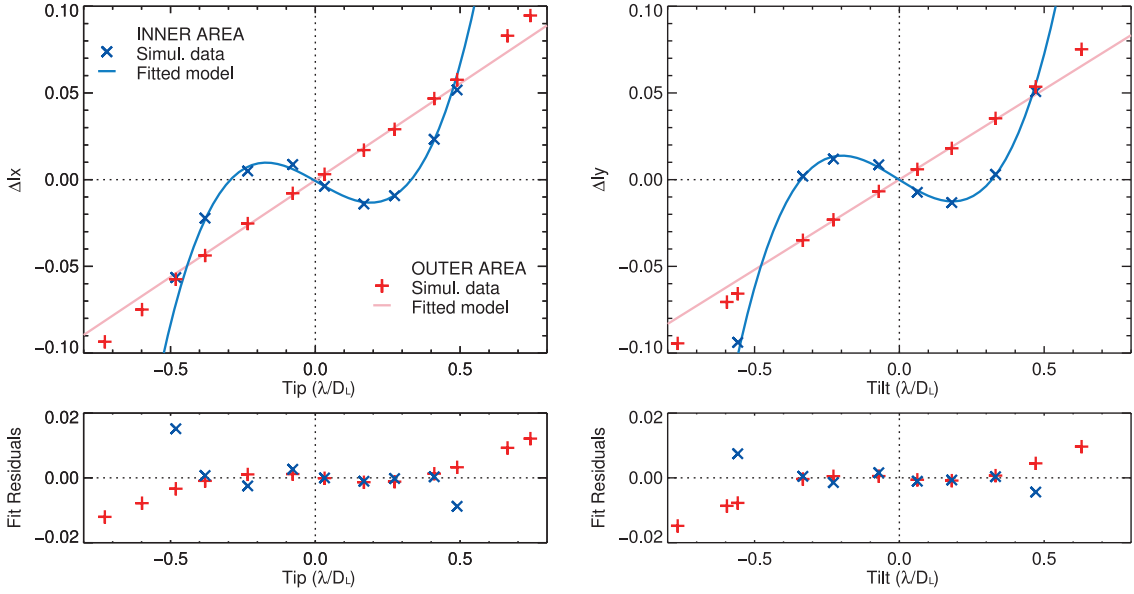} 
\caption{Same as Fig.~\ref{fig:exp_calib1} with data sets simulated for the same values of tip-tilt as the experimental data (including the offset in the orthogonal direction, as plotted in Fig.~\ref{fig:exp_true_tiptilt}), instead of actual measurements.}
\label{fig:exp_calib2}
\end{figure*}

\begin{table*}
\caption{Parameter values for the best fit models as defined by Eq.~\ref{eq:model}. The results are reported for the on-sky data sets (see Fig.~\ref{fig:exp_calib1}) as well as for the data sets simulated with the same tip-tilt sampling (see Fig.~\ref{fig:exp_calib2}). The ideal model corresponds to the result of the fit performed on the simulated model shown in Fig.~\ref{fig:ideal_model}, including a much larger number of simulated data points (not biased by the sampling or by the offset in the orthogonal direction of the applied tip/tilt). Values below $10^{-3}$ are considered insignificant and marked as 0.}
\label{tab:model_params}
\centering
\begin{tabular}{c|cccc|cccccccc}
\hline \hline
& \multicolumn{4}{c|}{Outer area (linear model)} &  \multicolumn{8}{c}{Inner area (cubic model)} \\
data set & $a$ & $\sigma_a$ & $b$ & $\sigma_b$ & $a$ & $\sigma_a$ & $b$ & $\sigma_b$ & $c$ & $\sigma_c$ & $x_0$ & $\sigma_{x_0}$\\
\hline
on-sky $x$ 		& 0.109	& $\pm$0.008 & 0.004	& $\pm$0.002	& 0.790	& $\pm$0.167	& -0.014	& $\pm$0.057	& -0.082	& $\pm$0.020	& 0.003	& $\pm$0.031 \\
simulated $x$ 	& 0.111	& $\pm$0.006	& 0	& $\pm$0.002	& 0.986	& $\pm$0.090	& -0.055	&$\pm$0.034	& -0.095	& $\pm$0.012	& 0	& $\pm$0.015\\
on-sky $y$ 		& 0.113 & $\pm$0.013 & 0.010 & $\pm$0.003 & 0.767& $\pm$0.045 & 0.128  & $\pm$0.015	& -0.084 & $\pm$0.005& -0.039& $\pm$0.007 \\
simulated $y$ 	& 0.104 & $\pm$0.004& 0 & $\pm$0.001& 0.976& $\pm$0.109 & 0.014& $\pm$0.032 & -0.106& $\pm$0.010 & -0.002 & $\pm$0.016 \\
ideal model		& 0.104& $\pm$0.007 & 0 	& $\pm$0.002 & 1.013 & $\pm$0.051& 0 & $\pm$0.019 & -0.146 & $\pm$0.007  & 0 & $\pm$0.009\\
\hline
\end{tabular}
\end{table*}

The relative position of the two components of the binary were first estimated in an unsaturated image, by fitting a Gaussian profile to each PSF. The vector connecting the position of the two star images is then used as a reference to estimate the true position of the primary star image with respect to the position of the secondary in the coronagraphic images. Given that observations are carried out in pupil tracking mode, the rotation of the position vector is also taken into account by correcting for the parallactic angle. The positions that have been probed during the tip-tilt scanning sequence are plotted in Fig.~\ref{fig:exp_true_tiptilt}, showing that the initial position that was assumed to be aligned with the vortex center was actually offset by about $-0.25\,$\lbdd in tip and tilt, due to an imperfect manual positioning at the beginning of the sequence. Based on these data, the off-axis peak transmission of the vortex coronagraph has been computed by integrating the flux in a disk of diameter $1\,$\lbdd centered on the actual star image position. As shown in Fig.~\ref{fig:exp_true_tiptilt}, these experimental results are in excellent agreement with the simulated curve, leading to an effective IWA of 125\,mas.

Figure \ref{fig:exp_calib1} shows the differential intensity measured in these images in the $x$ and $y$ directions as a function of the true tip and tilt, respectively. The data points corresponding to tip or tilt amplitude lower than 0.5\,\lbdd were fitted by polynomial functions expressed as
\begin{equation}
\begin{array}{l}
a x + b \mathrm{\text{, for the outer area, and}} \\
a (x-x_0)^3 + b (x-x_0)^2 + c(x-x_0) \mathrm{\text{, for the inner area}}.
\end{array}
\label{eq:model}
\end{equation}
All best fit parameters computed in the least-squares sense are reported in Table~\ref{tab:model_params}.

To validate the model, data sets were simulated with the same values of tip and tilt (including the offset in the respective orthogonal direction as shown in Fig.~\ref{fig:exp_true_tiptilt}). These simulated data were analyzed using the same fitting procedure. The results are shown in Fig.~\ref{fig:exp_calib2} and the best fit parameter values are reported in Table~\ref{tab:model_params}. The ideal model computed from simulations without offset (only pure tip or tilt was applied, as presented in Fig.~\ref{fig:ideal_model}) and with a finer sampling is also analyzed in the same way for comparison. The corresponding best fit parameters are also reported in the bottom line of Table~\ref{tab:model_params}.

As expected, the differential intensity measured in the outer area of the image is well approximated by a linear function, in particular for small tip-tilt amplitude (for tip-tilt amplitude $>\,0.5\,$\lbdd, the data points diverge from the linear model). The slopes are in agreement with the values predicted by simulations, but small global offsets are observed (non zero value for parameter $b$). This effect is stronger for the data corresponding to the $y$ direction. Possible causes for this effect will be discussed in the next subsection.

The differential intensity measured in the inner area shows the expected sign inversions around $\pm 0.4$\,\lbdd. The model parameter values fitted on the on-sky data are globally lower than the values that were expected from the simulations (from 14\% to 21\% for the first and third order terms) but are consistent with the simulations within error bars, except for the second order term in the $y$ direction. Besides, the coefficients for the second order terms were expected to be null according to the analytical model derived in \cite{Huby2015}, as confirmed by the simulations (see the best fit parameter value for the ideal model). As discussed in the next subsection, possible explanation for this behaviour includes the presence of additional asymmetric components in the image.

\begin{figure}
\centering
\includegraphics[height=0.88\linewidth, angle=270]{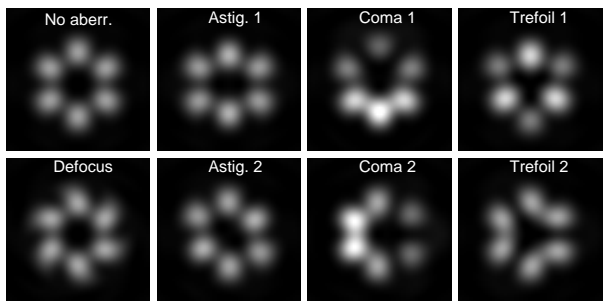}
\caption{Simulated images obtained with the vortex coronagraph on the Keck telescope, in presence of low order aberrations: 100\,nm rms defocus, 70\,nm rms for the others. All images are displayed with the same gray scale.}
\label{fig:aberr_img}
\end{figure}

\subsection{Sources of bias}

In its current state, the QACITS closed loop control tends to make the outer part of the coronagraphic image as symmetric as possible. However, this situation corresponds to the best centering of the star image on the mask only if the observed target is a point source, thus assuming that there is no bright asymmetric structure present in the very close vicinity of the central star (within $2.7\,$\lbdd or 240\,mas) and that the optical setup is not affected by other aberrations. While the former situation constitutes an intrinsic limitation of the QACITS estimator, bias due to optical imperfections can be mitigated to some extent, assuming that their cause is understood.

\subsubsection{Effect of low order aberrations}

\begin{figure}
\centering
\includegraphics[width=\linewidth]{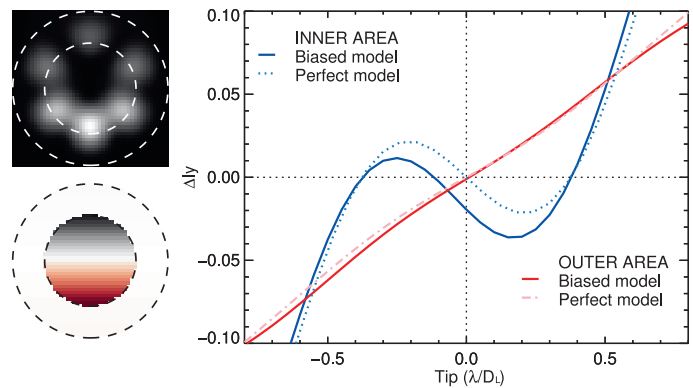} 
\caption{Effect of coma (70\,nm rms) on the QACITS model. On the left, the image shows the shape obtained when the star image is perfectly centered on the vortex mask, and the image below is the color-coded representation of the measured differential intensity in inner and outer regions of the image to emphasize the visual comparison of the asymmetry in both regions (the flux gradient in each area has an amplitude equal to the measured differential intensity).}
\label{fig:effect_coma}
\end{figure}

Low order aberrations have been investigated as sources of asymmetry in the final image. The resulting image shapes are shown in Fig.~\ref{fig:aberr_img}. While defocus and astigmatism have an effect on the shape of the image, they do not affect the central symmetry. On the other hand, coma and trefoil do. At the same aberration level, the effect of coma is almost one order of magnitude stronger than trefoil, and for that reason, its effect has been investigated in more detail.

The impact of coma on the QACITS model is shown in Fig.~\ref{fig:effect_coma}: the differential intensity measured in the inner area is significantly affected, while the outer differential intensity is mostly unchanged. Indeed, although it is not visually obvious, the asymmetry induced by coma is mostly concentrated in the inner area of the image, and the outer area is barely affected, as shown by the simulated image in Fig.~\ref{fig:effect_coma}. As a result, this aberration cannot explain the bias observed in the on-sky data.

\subsubsection{Effect of misalignment of the Lyot stop}
\label{sec:effect_lyotshift}

\begin{figure}
\centering
\includegraphics[width=\linewidth]{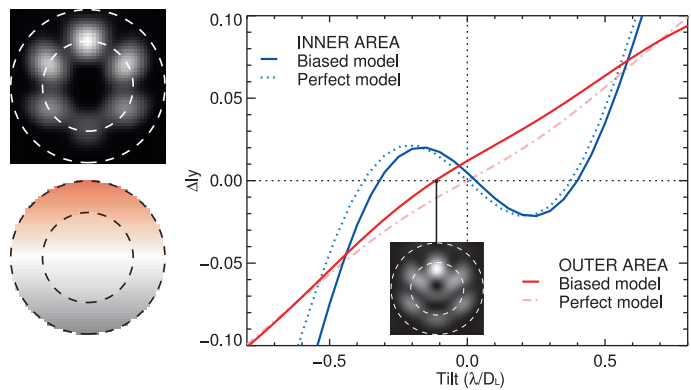} 
\caption{Same as Fig.~\ref{fig:effect_coma} in the case of a misalignment of the Lyot stop (shift of 4\% of the entrance pupil diameter, along the horizontal direction). The image obtained when the asymmetry in the outer area is compensated by tip-tilt is shown in the inset.}
\label{fig:effect_lyotshift}
\end{figure}

In case of a Lyot stop misalignment with respect to the entrance pupil, the coronagraphic image becomes asymmetric too, as illustrated in Fig.~\ref{fig:effect_lyotshift}. In particular, the differential intensity measured in the outer annulus is significantly affected, resulting in an offset of the QACITS model, as observed in the on-sky data. In practice, it means that the QACITS control loop, which is based on the outer estimator only, will converge towards a position that does not correspond to the best centered position. In the case of a 4\% shift relative to the entrance pupil diameter (as illustrated in Fig.\,\ref{fig:effect_lyotshift}), this bias has an amplitude of about $0.12\,$\lbdd.

\begin{figure}
\centering
\includegraphics[width=0.9\linewidth]{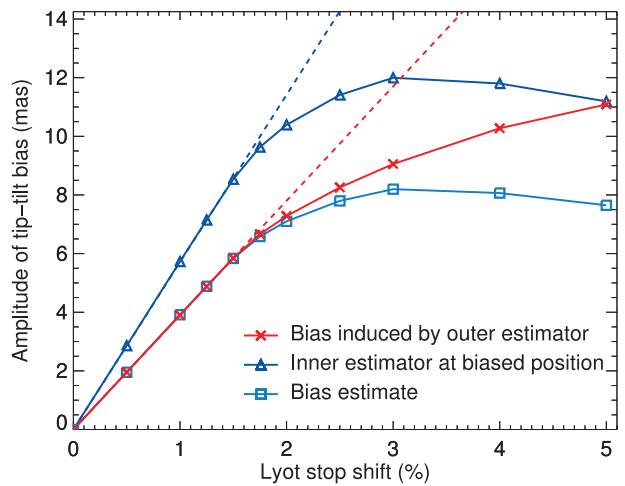}
\caption{Bias induced by the QACITS outer estimator, as a function of the Lyot stop shift, in units of percent of the telescope entrance pupil diameter (i.e. 1\% corresponds to 11\,cm at the scale of the telescope pupil). The red crosses show the bias induced by the outer estimator. The blue triangles are the inner estimator values as measured at the biased position. The light blue squares are the same estimates scaled by an adjustment factor of 0.7. This factor is computed as the slope ratio of the best linear model fitted on the actual bias and inner estimate at small Lyot shift amplitudes ($<2\%$), shown as dashed lines.}
\label{fig:bias_plot}
\end{figure}

The bias induced by the outer estimator depends on the amplitude of the Lyot stop shift with respect to the entrance pupil. This bias is estimated as the tip-tilt amplitude for which the differential intensity measured in the outer annulus cancels out. Simulation results reported in Fig.~\ref{fig:bias_plot} show that the bias increases linearly with the amplitude of the Lyot stop misalignment up to a 2\% shift. 

Based on these simulation results, we propose a method to estimate the amplitude of the bias affecting the outer estimator thanks to the inner estimator, which is less affected by the misalignment (as highlighted in Fig.\,\ref{fig:effect_lyotshift}): when the outer estimator loop has converged on the position where the outer differential flux is minimal, the inner estimator applied on that biased position leads to an estimate of the bias over-estimated by a factor $\sim1.5$ (see Fig.~\ref{fig:bias_plot}). Scaled by 70\%, this estimate can thus be used as a set point to correct for the bias of the outer estimator. For Lyot stop shifts smaller than 2.5\%, the residual bias when applying this method should be less than 0.01\,\lbdd, i.e. 1\,mas. The implementation of this additional step will result in a longer time dedicated to the optimization of the centering. Engineering time will be needed to implement and test this upgrade of the QACITS controller.

\section{Performance assessment}
\label{sec:perf}

First light of the vortex mode with the Keck/NIRC2 instrument was achieved in June 2015 \citep{Serabyn2017}. During this 3-night run, a preliminary version of the QACITS loop was closed on the second night. Another vortex run took place in October 2015, including one engineering night for the implementation of an improved version of the QACITS automated loop (as described in Sect.~\ref{sec:qacits_loop}) and for performing the calibration of the model (as reported in Sect.~\ref{sec:calib}). In this section, we present a comparison of the results obtained on data sets taken before (June 9th, 2015) and after (October 2015) the deployment of the optimized QACITS controller, highlighting the benefits of the control loop.

\subsection{Correction for the slow drift}
\label{sec:slow_drift}

In this subsection, we present a comparison of the QACITS estimator applied in post-monitoring on two data sets taken on the same target, HR8799, and under similar observing conditions: the target was observed close to transit (airmass $\sim1$), and the seeing was estimated to be $~0\farcs5$ during the hour preceding the acquisition sequence on June 9th (no seeing data during the sequence) and $~0\farcs7$ on average during the sequence on October 24th (CFHT DIMM seeing measured at 0.5\,$\mu$m).

\begin{figure}
\centering
\includegraphics[width=.97\linewidth]{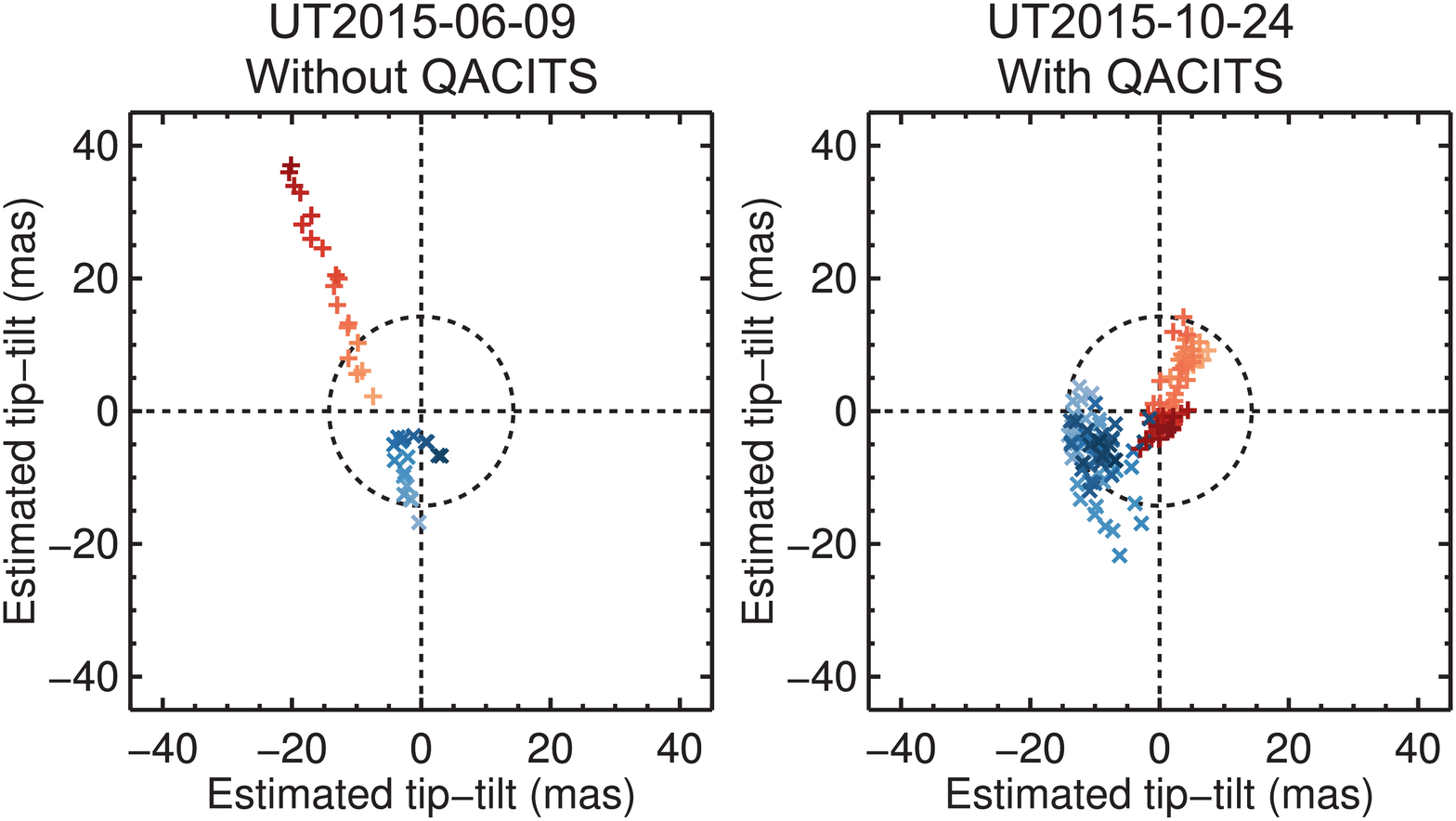}
\caption{QACITS position post-monitoring of the image of the target HR8799 onto the vortex mask in June (\textit{left}) and October 2015 (\textit{right}). The inner and outer estimators are plotted with blue crosses and red pluses, respectively. The dashed circles have a radius of 0.15\,\lbdd=13\,mas. The color shade of the points becomes darker with time. Every data point corresponds to 20\,s and 25\,s of integration time in June and October, respectively.}
\label{fig:loop_exple}
\end{figure}

During the night of June 9th, the QACITS controller was not yet operational and the star image was initially centered manually onto the vortex mask and maintained as well as possible by manually adjusting its position (every $\sim$ 10\,min) based on a visual assessment of the coronagraphic image shape. The data sequence taken on HR8799 is 15-min long. The results of the QACITS estimators applied in post-monitoring are shown in Fig.~\ref{fig:loop_exple} (left). For this particular data set, a clear drift is observed, at a rate of $\sim$2.7\,mas per minute. The inner estimator is in complete disagreement with the outer estimator because the tip-tilt amplitude rapidly reaches values outside the validity range of the linear approximation (typically $> 0.15\,$\lbdd i.e. 13\,mas). The way the modulus of the inner estimator decreases and then increases during the sequence is consistent with the expected behaviour around the change of sign of the derivative function (see Fig.~\ref{fig:ideal_model}), while the implemented estimator is based solely on the linear approximation.

For comparison, the same analysis has been performed on a sequence taken on the same target four months later, with an operational QACITS control loop. The sequence spans over 90\,min in total, including a gap without data due to the inability of tracking the star very close to zenith. The QACITS estimators displayed in Fig.~\ref{fig:loop_exple} (right) show a significant improvement in stability: the standard deviation of all outer estimates is 2.2\,mas and 5.1\,mas in the $x$ and $y$ directions, respectively. The larger dispersion observed along the $y$ axis indicates the probable direction of the drift that had to be corrected by the controller. The inner estimates are somewhat offset, with a mean amplitude of 11.7\,mas. Based on the simulation results assuming a misalignement of the Lyot mask with respect to the entrance pupil (Fig.~\ref{fig:bias_plot}), this amplitude can be interpreted as a shift of the Lyot stop of up to 2.7\% at the time of these observations.

\subsection{Pointing statistics}
\label{sec:pointing_stats}

\begin{figure}
\centering
\includegraphics[width=.97\linewidth]{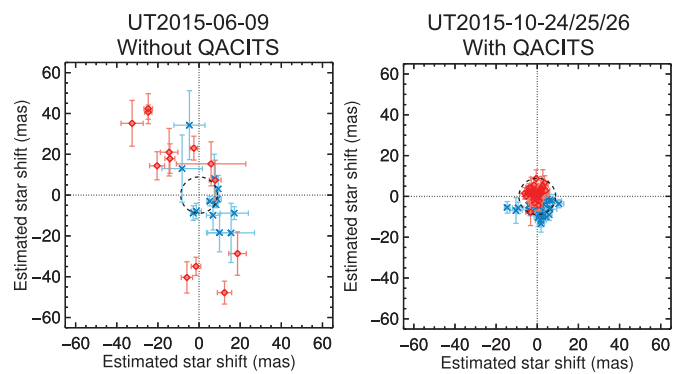}
\caption{On-sky results from the QACITS post-monitoring of data sets taken during the night of first light (\textit{left}) and on a later run in October with the control loop closed (\textit{right}). Every point represents the mean estimate of a sequence (comprising between 8 and 72 acquisition frames, 10 in average), with the error bars showing the standard deviation of the estimates during the sequence. The dashed circles have a radius of 0.1\,\lbdd=9\,mas.}
\label{fig:stats}
\end{figure}

The same post-monitoring procedure based on the QACITS estimators has been applied to all data sequences taken on the night of June 9th, 2015, and during three consecutive nights dedicated to science targets in October 2015. The mean outer and inner estimates of every sequence are plotted in Fig.~\ref{fig:stats}. The comparison of the results with and without the QACITS controller is quite explicit: on the June night, the dispersion of the mean outer estimators is 16\,mas and 32\,mas in the $x$ and $y$ directions, respectively, while it is reduced to 2.3\,mas and 2.6\,mas, respectively, during the three observing nights in October. The average standard deviation within every sequence is also indicative of the improved stability, as it is reduced from 4.1\,mas and 7.9\,mas on June 9th (in the $x$ and $y$ directions, respectively), down to 2.0\,mas and 2.8\,mas over the three nights in October.

Additionally, for the data taken without the automated centering of the star, the tip-tilt amplitude for the outer estimator reaches 33\,mas (0.37\,\lbdd) on average, which roughly corresponds to the amplitude for which the flux asymmetry in the inner disk changes sign (see the model curve in Fig.~\ref{fig:ideal_model}). In other words, around this particular tip-tilt value, there is almost no asymmetry visible in the inner region of the image, which appears as a symmetric bright doughnut (see Fig.~\ref{fig:simulated_img}). As a result, the observer can be easily tricked by this apparent symmetry, and can consider that the star image is centered onto the vortex mask, while it is actually offset by about 30\,mas from the vortex mask center.

In contrast, on the October nights, the mean outer estimators are all significantly closer to zero (as expected since the control loop was based on the outer estimator only), but the mean inner estimators still show a systematic offset, of amplitude 6.4\,mas on average. Assuming that this effect is due to a misalignment of the Lyot mask, and based on simulation results (Fig.~\ref{fig:bias_plot}), this amplitude indicates that the outer estimator is affected by a bias of 4.4\,mas on average (corresponding to a shift of the Lyot stop of $\sim$1.1\%, which lies within the specifications of the alignment accuracy). It has to be noted, though, that this systematic error is relatively constant over the three nights, with a preferred direction. This means that both the science and reference targets are affected in the same way to some extent, and the impact on differential imaging techniques is therefore limited \citep{Mawet2017, Serabyn2017}.

\section{Conclusion and prospects}
\label{sec:conclusion}

The QACITS controller has been successfully implemented on the newly commissioned vortex mode of the Keck/NIRC2 instrument. The benefits of this automated control are multiple: 
\begin{itemize}
\item The observation sequence is fully automated, including taking calibration frames, initial centering, and stabilization of the star during the observation, making the vortex mode sufficiently user friendly to be offered to the community (in shared-risk mode since 2016B).
\item Pointing quality is not observer-dependent, and in particular, the pitfall induced by the apparent symmetry of the coronagraphic shape when the star is offset by $\sim$ 30\,mas is avoided.
\item Pointing stability of 2.4\,mas rms is achieved on average, with the control loop running at a frequency of about 0.02\,Hz, thus correcting for low frequency drifts.
\item Pointing accuracy of 4.5\,mas is achieved on average. This accuracy is currently limited by systematic errors induced by a probable misalignment of the Lyot stop with respect to the entrance pupil. Still, the final average accuracy provides an improvement of a factor 7 over the accuracy achieved manually, and a method to reduce this bias down to the 1\,mas level is proposed.
\end{itemize}

Given the success and benefits of the QACITS controller, efforts are currently ongoing to develop the same kind of control loop on other operational infrared vortex coronagraphs, namely on the VLT/NACO, LBT/LMIRCam and VLT/VISIR instruments. The QACITS algorithm is also under study for implementation on the future mid-infrared ELT/METIS instrument \citep{Brandl2014}, which includes a vortex coronagraph in its baseline design. It can be noted that the implementation at other wavelengths is straightforward: since the tip-tilt amplitude is measured in units of \lbdd, the model is not dependent on wavelength. As a matter of fact, the very first laboratory tests with a non obstructed pupil were performed in K band with the vortex coronagraph on PHARO at Palomar Observatory \citep{Mawet2010}, and observations at M band were recently successfully carried out with the vortex coronagraph at Keck/NIRC2. Even more generally, the basic principle of the method, initially proposed by \cite{Mas2012} for the Four Quadrant Phase Mask with a non obstructed pupil, may potentially be adapted not only to the vortex coronagraph but also to other small IWA coronagraphs based on a focal plane mask (e.g. the Dual-Zone Phase Mask \citep[DZPM,][]{Soummer2003} or the Phase Induced Amplitude Apodizer \citep[PIAA,][]{Guyon2003}). The major adjustment concerns the characterization of the image behaviour in presence of tip-tilt and more specifically the definition of the underlying model for the measured differential flux, which may be derived either analytically and/or empirically, based on experimental calibration.

Lastly, we intend to use the same kind of method combining the inner and outer estimators in the data processing. Post-monitoring of the data using QACITS can indeed provide a means to perform frame selection based on a centering quality criterion, and not only on a flux criterion subject to seeing conditions. Besides, the estimate of the true star image position behind the coronagraphic mask allows a better registration of the frames and can thus potentially increase the signal to noise ratio of planets present in the rotating field.

\begin{acknowledgements}
The research leading to these results has received funding from the European Research Council under the European Union's Seventh Framework Programme (ERC Grant Agreement n. 337569) and from the French Community of Belgium through an ARC grant for Concerted Research Action.
\end{acknowledgements}

\bibliographystyle{aa}
\bibliography{Huby_2016_QACITS.bib}

\end{document}